\documentclass[aps,pra,twocolumn,amsmath,amssymb,groupedaddress,longbibliography]{revtex4-2}
\usepackage{graphicx}
\usepackage{dcolumn}
\usepackage{bm}
\usepackage[pdfstartview=FitH, CJKbookmarks=true, bookmarksnumbered=true, bookmarksopen=true, colorlinks=true, pdfborder=001, citecolor=blue, linkcolor=blue, linktocpage=true] {hyperref}

\newcommand{\nn}{\nonumber}

\begin{document}

\preprint{APS/123-QED}
\title{Edge states in quantum spin chains: the interplay among interaction, gradient magnetic field and Floquet engineering}

\author{Wenjie Liu $^{1,2}$}

\author{Bo Zhu $^{1,3}$}

\author{Li Zhang $^{1}$}

\author{Yongguan Ke $^{1}$}
\altaffiliation{Email: keyg@mail2.sysu.edu.cn}

\author{Chaohong Lee $^{1,2}$}
\altaffiliation{Email: lichaoh2@mail.sysu.edu.cn}

\affiliation{$^{1}$Guangdong Provincial Key Laboratory of Quantum Metrology and Sensing $\&$ School of Physics and Astronomy, Sun Yat-Sen University (Zhuhai Campus), Zhuhai 519082, China}

\affiliation{$^{2}$State Key Laboratory of Optoelectronic Materials and Technologies, Sun Yat-Sen University (Guangzhou Campus), Guangzhou 510275, China}

\affiliation{$^{3}$Institute of Mathematics and Physics, Central South University of Forestry and Technology, Changsha 410004, China}

\date{\today}

\begin{abstract}

We explore the edge defects induced by spin-spin interaction in a finite paradigmatic Heisenberg spin chain.
By introducing a gradient magnetic field and a periodically-modulated spin-exchange strength, the resonance between modulation frequency and magnetic field gradient can also induce asymmetric defects at the edges, dubbed as Floquet-Wannier-Zeeman edge defects.
The interplay between these two types of edge defects allows us to manipulate the magnon edge states from an isolated band into a continuum one.
In the high-frequency regime, we analytically derive effective models to interpret the formation mechanisms for edge states by employing the multiscale perturbation analysis.
Our study offers new insights to understand and control the magnon edge states governed by the interplay between edge defects induced by the spin-spin interaction and the Floquet-Wannier-Zeeman manipulation.

\end{abstract}

\maketitle

\section{Introduction\label{Sec1}}

Quantum spin chains serve as archetypes for understanding a variety of problems, such as, quantum magnetism~\cite{USchollwock2004}, quantum phase transition~\cite{SSachdev1999}, statistical mechanics~\cite{DWRobinson1967}, and so on.
Experimental platforms involving ultracold atoms~\cite{MAnderlini2007,JSimon2011,JStruck2011,TFukuhara20131,TFukuhara20132}, ultracold ions~\cite{AFriedenauer2008,KKim2009,EEEdwards2010,KKim2011}, superconducting circuits~\cite{JQYou2010,ULasHeras2014,YSalathe2015,RBarends2016,GWendin2017}, and nuclear magnetic resonance systems~\cite{XPeng2005,XPeng2010,ZLi2014} provide unprecedented opportunities for simulating quantum spin chains.
Dating back to Bethe's seminal work~\cite{HBethe1931}, magnons were predicted as collective excitations around ferromagnetic ground states in a quantum spin chain.
Due to strong spin-spin interactions, multiple magnons form bound states which were observed in solid-state materials~\cite{MDate1966,JBTorrance1969,RHoogerbeets1984}.
However, until the last decade nonequilibrium dynamics of magnon and its bound states were observed in optical lattices of ultracold atoms~\cite{TFukuhara20131,TFukuhara20132}.

The bulk properties of spin chains are usually investigated via systems under periodic boundary conditions.
Besides to the bulk states, edge states of spin chain are also very important in potential applications.
Edge states may arise as a result of symmetry-protected topological order in a generalized Heisenberg chain now known as AKLT model~\cite{IAffleck1987}.
However, such a model is far from accessible in experiments.
Edge states may appear due to elaborate design of spin exchange.
State transfer between two endpoints of a spin chain has been proposed by designing space-dependent spin-spin exchange~\cite{FMei2018,XLi2018} as well as introducing off-diagonal impurities~\cite{AZwick2011}.
Moreover, the artificially added defects offer a flexible mean to generate localized states and manipulate the entanglement in a quantum spin chain~\cite{TJGApollaro2006,FPlastina2007}.
In a fully polarized open spin chain, a spin flipped in one endpoint will stay localized in the time evolution, that is the so-called edge-locking effects.
The edge-locking phenomena have been revealed via the Bethe ansatz~\cite{VAlba2013} and the spectral structure analysis~\cite{MHaque2010,ASharma2014}.
However, correlation properties of the interaction-induced edge states are still unclear.
Furthermore, one may ask how to tailor the spectral structure, correlation properties, and quantum dynamics of the interaction-induced magnon edge states.

Floquet engineering and gradient magnetic field serve as important tools to manipulate the energy spectrum and quantum dynamics~\cite{GDellaValle2014,HZhong2017,WLiu1911}.
In a tilted lattice, a particle will change from Bloch oscillations~\cite{WLiu2019} to coherent delocalization by introducing periodic resonant drivings that smooth the bias potential, which have been applied for precision gravity measurements~\cite{VVIvanov2008,MGTarallo2012}.
Different from the coherent delocalization in the bulk, edge localization exists due to the appearance of effective edge defects in a driven tilted defect-free lattice~\cite{BZhu2020}.
Importantly, the energy of the Floquet-surface edge states can be tuned into and out of the continuum spectrum by changing the modulation parameters.
Stimulated by the Floquet-surface edge states in a driven tilted lattice, we engineer another type of effective defects which are called as Floquet-Wannier-Zeeman edge defects due to the joint effect of both the gradient magnetic field and periodically-modulated spin-exchange strength in the Floquet-Wannier-Zeeman spin chain.
In the absence of the periodically-modulated spin-exchange strength, the Floquet-Wannier-Zeeman edge defects can change to Wannier-Zeeman edge defects of spin chains subject to a gradient magnetic field.
Naturally, a meaningful question to ask is how the interplay between the interaction-induced edge defects and Floquet-Wannier-Zeeman edge defects affects the magnon edge states.

In this paper, we investigate the emergent magnon edge states in a finite simple Heisenberg spin chain which can be further tuned by introducing the gradient magnetic field and periodically-modulated spin-exchange strength.
In the absence of the gradient magnetic field and periodically-modulated spin-exchange strength, defects with energy determined by interaction strength localize a single magnon in the endpoints and repulse the magnon in the bulk.
Based on the correlation properties of states, we distinguish and name different isolated bands of energy spectrum in the presence of interaction-induced edge defects.
Interestingly, we find two degenerate bound-magnon edge states via the correlation analysis, that is, one magnon is trapped in a endpoint and the other bounds together.
This is because spin-spin interactions create edge defects and make the two magnons bound together at the defects.
According to the relation between interaction strength and edge defects, we are able to predict the positions of different kinds of states in the energy spectrum.
Floquet-Wannier-Zeeman edge defects appear in the so-called Floquet-Wannier-Zeeman spin chain which could be realized by introducing a gradient magnetic field and a periodically-modulated spin-exchange strength.
Two-magnon edge states in the continuum (EIC) are observed under the influence of the Floquet-Wannier-Zeeman edge defects.
Under the interplay between the two types of defects, the interaction-induced bound-magnon edge states can be tuned into the continuum band which can be identified by the inverse participation ratios (IPRs).
While both the two endpoints have defects induced by interaction or Floquet-Wannier-Zeeman manipulation, the edge defect at one endpoint can be tuned to vanish by the interplay between the two types of defects.
These results can be explained via static effective models obtained by employing the multiscale perturbation analysis.

This rest of paper is organized as follows.
In Sec.~\ref{Sec2}, we clarify interaction-induced edge defects in a finite spin chain.
In Sec.~\ref{Sec3}, we study Floquet-Wannier-Zeeman edge defects and analyze the interplay between these two types of defects based on the multiscale perturbation analysis.
In Sec.~\ref{Sec4}, we give a brief summary.

\section{Interaction-induced edge defects \label{Sec2}}
In the absence of the gradient magnetic field and periodically-modulated spin-exchange strength, we consider a typical spin-1/2 Heisenberg $XXZ$ chain described by the Hamiltonian
\begin{eqnarray}\label{Heisenberg}
\hat{H}_s=\sum\limits_{l=1}^{L-1} \big(\frac{J_0}{2}\hat{S}^+_l\hat{S}^-_{l+1}+{\rm H.c.}+\Delta\hat{S}^z_l\hat{S}^z_{l+1}\big).  \\ \nn
\end{eqnarray}
Here, $\hat{S}^i_l (i=x,y,z)$ are spin-1/2 operators and $\hat{S}^{\pm}_l=\hat{S}^x_l\pm i\hat{S}^y_l$ are spin raising and lowing operators at the $l$-th site.
It has been demonstrated that ultracold atoms in optical lattices offer a powerful platform to simulate quantum spin chains.
In the deep Mott-insulator regime, the quantum Heisenberg $XXZ$ chain~\eqref{Heisenberg} can be derived by using the second-order degenerate perturbation theory for ultracold two-component atoms in optical lattices~\cite{MTakahashi1997,Lee2004,LiZhang2019}.
Such a quantum Heisenberg $XXZ$ chain~\eqref{Heisenberg} has been experimentally realized using ultracold atoms~\cite{TFukuhara20131,TFukuhara20132}, in which magnon excitations, magnon bound states and their dynamics have been observed.
$\Delta$ is the spin-spin interaction which can be tuned via Feshbach resonance~\cite{AWidera2004,CGross2010}.
$J_0$ as the spin-exchange strength is set as unity (i.e., $J_0=\hbar=1$).

The eigenvalues and eigenstates of this typical spin-1/2 Heisenberg $XXZ$ chain can be solved analytically by using the Bethe ansatz~\cite{HBethe1931}.
For $\Delta>1$, the ground state of $XXZ$ chain has a N\'{e}el order along the $z$ direction.
For $\Delta<-1$, the phase is characterized by a ferromagnetic order along the $z$ direction.
In the region $-1<\Delta\leq1$, the system is in the spin-liquid phase.
Since the ground state properties are well known, we focus on a single- and two-magnon excitations over a ferromagnetic ground state $|\textbf{0}\rangle=|\downarrow\downarrow\downarrow...\downarrow\rangle$.
Not limited to the properties in the bulk, we turn our attention to the phenomena occurring on the edges of such spin chain.
The edge-localization phenomena in the open-boundary Heisenberg $XXZ$ chain have been investigated by employing the Bethe ansatz approach~\cite{VAlba2013} or the spectral structure analysis~\cite{MHaque2010,ASharma2014}.
Nevertheless, we provide a direct explanation of edge localization from the perspective of edge defects named as interaction-induced defects, which can also be applied to Floquet spin chains.

\subsection{Single-magnon excitation \label{Sec21}}

\begin{figure}[htp]
\begin{center}
\includegraphics[width=0.45\textwidth]{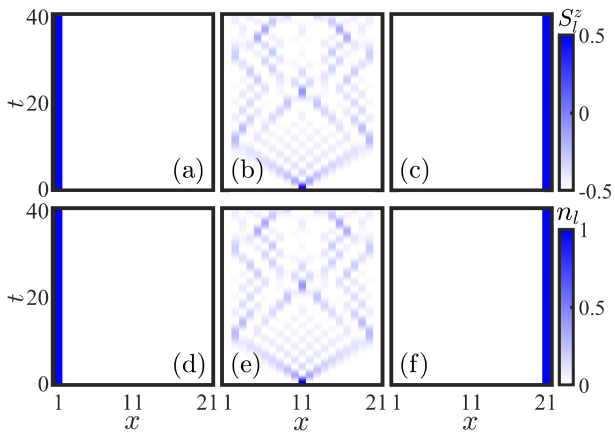}
\end{center}
\caption{(Color online)
Dynamical evolution for single-magnon excitation under the open boundary condition.
The time evolution of spin magnetization calculated with the time-evolving block decimation algorithm for $\Delta=20$ with initially flipped spin at the left endpoint (a), center (b) and right endpoint (c).
The time evolution of magnon distribution calculated with exact diagonalization for $\Delta=20$ with initially prepared magnon at left endpoint (d), center (e) and right endpoint (f). The length of chain is $L=21$.
The coordinate unit is set as the lattice constant and the time unit is set as $2\pi/J_0$.}
\label{fig:onemagnondynamics}
\end{figure}

For an initially flipped spin over the fully ferromagnetic state with all spins downward, we track out the time evolution of spin magnetization
\begin{eqnarray}
S^z_{l}(t)=\langle\psi(t)|\hat{S}^z_{l}|\psi(t)\rangle
\end{eqnarray}
by time-evolving block decimation algorithm with $L=21$ and $\Delta=20$.
For open boundary condition, an interesting phenomenon is that dynamical localization appears when the initial spin is flipped at the left or right endpoint [see Figs.~\ref{fig:onemagnondynamics}(a) and~\ref{fig:onemagnondynamics}(c)];
Another relative phenomenon is the so-called repulsion effect, that is, the spin transport is repelled from reaching the boundary sites and the distribution at sites $1$ and $L$ remains spin down all the time when the spin is flipped at the center of the spin chain [see Fig.~\ref{fig:onemagnondynamics}(b)].
However, these abnormal phenomena disappear for $\Delta=0$ and are enhanced by increasing $\Delta$.

To better understand these phenomena, we transfer to the language of magnon excitation which is obtained by flipping spins over the ferromagnetic ground state $|\downarrow\downarrow\downarrow...\downarrow\rangle$.
Magnon as the collective excitation around the ferromagnetic ground state follows the commutation relations of hard-core bosons.
By employing the mapping $\left|\downarrow\right\rangle\leftrightarrow \left|0\right\rangle, \left|\uparrow\right\rangle\leftrightarrow \left|1\right\rangle, \hat{S}^+_l\leftrightarrow\hat{a}^{\dag}_l, \hat{S}^-_l\leftrightarrow\hat{a}_l$, and $  \hat{S}^z_l\leftrightarrow\hat{n}_l-\frac{1}{2}$,
it gives insights into the properties of the quantum magnetic systems from the perspective of magnon excitations,
i.e.,
\begin{eqnarray}\label{magnonmodel} \nn
\hat{H}_m=&\sum\limits_{l=1}^{L-1} \left(\frac{J_0}{2}\hat{a}^{\dag}_l\hat{a}_{l+1}+{\rm H.c.}+\Delta\hat{n}_l\hat{n}_{l+1}\right)-\frac{\Delta}{2}\hat{n}_1-\frac{\Delta}{2}\hat{n}_L. \\
\end{eqnarray}
$\hat{a}_{l}^{\dag}$ ($\hat{a}_l$) creates (annihilates) a magnon at the $l$-th site and
$\hat{n}_l=\hat{a}_{l}^{\dag}\hat{a}_l$ is the magnon number operator.
Remarkably, the nearest-neighbor interaction, corresponding to the spin-spin interaction in the spin chain~\eqref{Heisenberg}, introduces two defects with the same value and sign on the two endpoints.
It is reasonable to name it as interaction-induced defects which can explain the dynamical localization and repulsion effects, that is,
a large energy gap between each endpoint and bulk strongly suppress the coupling between each endpoint and bulk.
The interaction-induced edge defects vanish when the interaction is turn off and increase with the interaction strength.
Due to $[\hat{H}_s,\hat{S}^z]=0$ with $\hat{S}^z=\sum_l\hat{S}^z_l$, the total spin magnetization along the $z$ direction is conserved.
The corresponding magnon number $\hat{N}=\sum_l\hat{n}_l$ is also conserved. It means that subspaces with different numbers of magnons are decoupled.
In the $N$-magnon basis $\{\hat{a}^\dag_{l_1}\hat{a}^\dag_{l_2}...\hat{a}^\dag_{l_N}|\textbf{0}\rangle\}$ with $1\leq l_1<l_2<...<l_N\leq L$,
the arbitrary states of system can be expanded as $|\psi\rangle=\sum_{l_1<l_2<...<l_N}\psi_{l_1l_2...l_N}|l_1l_2...l_N\rangle$ with the probability amplitude $\psi_{l_1l_2...l_N}=\langle\textbf{0}|\hat{a}_{l_N}...\hat{a}_{l_2}\hat{a}_{l_1}|\psi\rangle$.

For a comparison, at $t=0$ we respectively prepare a single magnon at the left endpoint, the center and the right endpoint, and the dynamics of magnon density distribution $n_{l}(t)=\langle\psi(t)|\hat{n}_{l}|\psi(t)\rangle$ is obtained by exact diagonalization of the Hamiltonian~\eqref{magnonmodel} with 21 lattice sites, as shown in Figs.~\ref{fig:onemagnondynamics}(d)-\ref{fig:onemagnondynamics}(e).
It is clear that the magnon dynamics is in full agreement with the spin transport in Figs.~\ref{fig:onemagnondynamics}(a)-\ref{fig:onemagnondynamics}(c) under the same parameters.
Hence, in the following context we frequently use the language of magnon excitations to simplify our analysis.

\subsection{Two-magnon excitations \label{Sec22}}

\begin{figure}[htp]
\begin{center}
\includegraphics[width=0.45\textwidth]{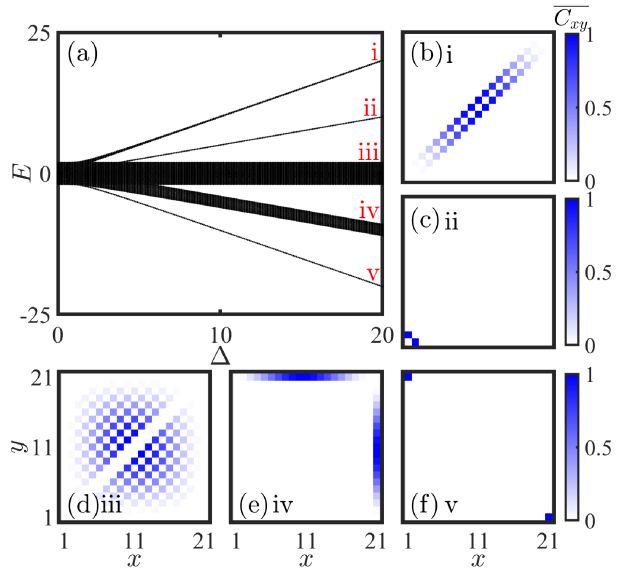}
\end{center}
\caption{(Color online)
(a) Energy spectrum as a function of nearest-neighbor interaction for $L=21$.
The subsets of spectrum from top to bottom correspond to (i) bound magnon pair, (ii) bound-magnon edge state, (iii) independent magnons, (iv) one-magnon edge state and (v) two-magnon edge state.
(b)-(f) The normalized two-magnon correlations $\overline{C_{xy}}=C_{xy}/C^{max}_{xy}$ of the states for $\Delta=20$ labeled with (i), (ii), (iii), (iv) and (v) in (a).
The energy unit is set as $J_0=\hbar=1$ and the coordinate unit is set as the lattice constant.}
\label{fig:twomagnonspectrum}
\end{figure}

Then we turn to explore the two-magnon problem.
We calculate the energy spectrum as a function of the nearest-neighbor interaction regarding the Hamiltonian~\eqref{magnonmodel}, see Fig.~\ref{fig:twomagnonspectrum}(a).
The chain length is $L=21$.
In comparison to the noninteracting case, the energy spectrum becomes dramatically rich as the appearance of more isolated bands.
For $\Delta=20$, we respectively pick one state from five bands labeled with (i), (ii), (iii), (iv), (v) and analyze their correlation properties in Figs.~\ref{fig:twomagnonspectrum}(b)-\ref{fig:twomagnonspectrum}(f).
The two-magnon correlation is defined as
\begin{equation}
C_{xy}(t)=\langle\psi(t)|\hat{a}^{\dag}_{x}\hat{a}^{\dag}_{y} \hat{a}_{y}\hat{a}_{x}|\psi(t)\rangle.
\end{equation}
$x$ and $y$ denote the lattice sites and span from $1$ to $L$.
The two-magnon correlations at two specific lines $x=y\pm d$ in the $(x,y)$ plane characterize the two-magnon bound states, where $d$ relies on the specific two-magnon interactions.
There exists a connection between the spin correlation $S_{xy}(t)=\langle\psi(t)|\hat{S}^{z}_{x}\hat{S}^{z}_{y}|\psi(t)\rangle$ with the two-magnon correlation $C_{xy}(t)=\langle\psi(t)|\hat{S}^{+}_{x}\hat{S}^{+}_{y} \hat{S}^{-}_{y}\hat{S}^{-}_{x}|\psi(t)\rangle$~\cite{WLiu2019} yielding
\begin{eqnarray}
S_{xy}=\left\{
\begin{aligned}
C_{xy}-\frac{1}{2}S^z_{x}&-\frac{1}{2}S^z_{y}-\frac{1}{4} &\text{if}\  x\ne y, \\
&\frac{1}{4} &\text{if}\  x=y.
\end{aligned}
\right.
\end{eqnarray}
Different types of states are exhibited via the two-magnon correlations: bound magnon pair with $E\approx\Delta$ [Fig.~\ref{fig:twomagnonspectrum}(b)], bound-magnon edge state with $E\approx\Delta/2$ [Fig.~\ref{fig:twomagnonspectrum}(c)], independent magnons with $-2J_0\leq E \leq2J_0$ [Fig.~\ref{fig:twomagnonspectrum}(d)], one-magnon edge state with $E\approx-\Delta/2$ [Fig.~\ref{fig:twomagnonspectrum}(e)] and two-magnon edge state with $E\approx-\Delta$ [Fig.~\ref{fig:twomagnonspectrum}(f)].
We also observe bound-magnon edge state localized at the right edge with the same energy with that in Fig.~\ref{fig:twomagnonspectrum}(c),
and one-magnon edge state for the magnon distributes at the left edge with the same energy with that in Fig.~\ref{fig:twomagnonspectrum}(e).
As the nearest-neighbor interaction increases, the energy spectrum is gradually separated from one band into five bands whose slopes regarding the interaction nearly yield 1 [(i)], 0.5 [(ii)], 0 [(iii)], -0.5 [(iv)] and -1 [(v)].
The existence of bound-magnon edge states indicates that the two-site interaction not only facilitates the formation of magnon bound pair, but also attracts this pair locating on one endpoint of the lattice.
Apart from localization and repulsion phenomena of a single magnon, the bound-magnon edge state, one- and two-magnon edge states can be also understood as a consequence of the interaction-induced defects at the outermost sites.

\section{Floquet-Wannier-Zeeman manipulation  \label{Sec3}}

In this section, after introducing the gradient magnetic field and periodically-modulated spin-exchange strength, we first show edge localizations in the absence of interaction of the Floquet-Wannier-Zeeman spin chain.
Then, it is natural and interesting to explore the interplay between the interaction-induced edge defects and Floquet-Wannier-Zeeman edge defects.
We find asymmetric transports and bound-magnon edge states in the continuum (BEIC), as joint effects of interaction and Floquet-Wannier-Zeeman manipulation.
To obtain a deeper understanding of these phenomena, we analytically derive static effective models based on multiscale perturbation analysis from single- to two-magnon systems.
These static effective models not only tell how to tune symmetric magnon transports into asymmetric ones, but also perfectly explain the
appearance of two-magnon EIC and BEIC.

\subsection{Floquet-Wannier-Zeeman spin chain \label{Sec30}}

By introducing a gradient magnetic field $B$ and a periodically-modulated spin-exchange strength $J(t)$ into the Hamiltonian~\eqref{Heisenberg}, we obtain a Floquet-Wannier-Zeeman spin chain
\begin{eqnarray}\label{originalspinchain}\nn
  \hat{\mathcal{H}}_s=\sum_{l=1}^{L-1}\big[J(t)\hat{S}^+_l\hat{S}^-_{l+1}+\rm{H.c.}+\Delta\it{\hat{S}^z_l\hat{S}^z_{l+\rm{1}}+lB\hat{S}^z_l} \big].\\
\end{eqnarray}
The experimental probability of the Floquet-Wannier-Zeeman spin chain~\eqref{originalspinchain} has been discussed by using ultracold atoms in optical lattices~\cite{WLiu1911}.
Different from the bulk properties in~\cite{WLiu1911}, we focus on the edge states.
The edges could in principle be realized by applying a sharp box potential into the optical lattices~\cite{ALGaunt102004062013}.
The corresponding model of magnons in a driven and tilted lattice is given by
\begin{eqnarray}\label{magnonmodel2} \nn
\hat{\mathcal{H}}_m=&\sum\limits_{l=1}^{L-1} \left[J(t)\hat{a}^{\dag}_l\hat{a}_{l+1}+{\rm H.c.}+\Delta\hat{n}_l\hat{n}_{l+1}\right] \\
&-\frac{\Delta}{2}\hat{n}_1-\frac{\Delta}{2}\hat{n}_L+\sum\limits_{l=1}^{L} Bl\hat{n}_l
\end{eqnarray}
with $J(t)=[J_0+J_1 \cos(\omega t)]/2$.
$J_1$ and $\omega$ serve as modulation amplitude and modulation frequency, respectively.

Referring to the works~\cite{ABuchleitner2003,ARKolovsky2003}, we apply a unitary operator $\hat{U}=\exp({i\sum_l lBt\hat{n}_l})$ to transform the Hamiltonian~\eqref{magnonmodel2} into a rotating frame according to $\hat{\mathcal{H}}^{rot}_m=\hat{U}\hat{\mathcal{H}}_m\hat{U}^{{\dag}}-i\hat{U}\frac{\partial}{\partial t}\hat{U}^{{\dag}}$.
We obtain the Floquet magnon Hamiltonian in the rotating frame
\begin{eqnarray}\label{FloquetH}
\hat{\mathcal{H}}^{rot}_m&=&\sum\limits_{l=1}^{L-1}\big[\mathcal{J}(t)\hat{a}^{\dag}_l\hat{a}_{l+1}+{\rm H.c.}+\Delta \hat n_l\hat n_{l+1}\big] \nonumber \\
&-&\frac{\Delta}{2}\hat{n}_1-\frac{\Delta}{2}\hat{n}_L
\end{eqnarray}
with $\mathcal{J}(t)=M_0 e^{i(\omega-B) t}+M_1 e^{-iB t}+M_2 e^{-i(\omega+B) t}$.
Constrained by the realization of the Floquet magnon Hamiltonian~\eqref{FloquetH}, we have $M_0=J_1/4$, $M_1=J_0/2$ and $M_2=J_1/4$.
At a resonant condition $\omega=B$, the driving frequency matches and smooths the potential bias between neighboring sites~\cite{VVIvanov2008,MGTarallo2012,NGoldman2015}.
In this paper, we focus on the resonant driving with $\omega=B$ to obtain $\mathcal{J}(t)=M_0 +M_1 e^{-i\omega t}+M_2 e^{-i2\omega t}$.
Clearly, the resonant driving yields a two-color modulation and satisfies a discrete time translation invariance $\hat{\mathcal{H}}^{rot}_m(t+T)=\hat{\mathcal{H}}^{rot}_m(t)$ with a period $T=2\pi/\omega$, which can be analyzed by the Floquet theory.
At each cycle, the dynamics is governed by a time-ordering operator,
\begin{eqnarray}
\hat{U}_T=\hat{\mathcal{T}}e^{-i\int^T_0\hat{\mathcal{H}}^{rot}_m(t)dt}\equiv e^{-i\hat{H}_F T}
\end{eqnarray}
with an effective time-independent Hamiltonian
\begin{eqnarray} \label{staticHamitonian}
\hat{H}_F=\frac{i}{T}\log\hat{U}_T.
\end{eqnarray}
One can obtain quasienergies and Floquet states by solving the eigenequation
$\hat{H}_F|u_n\rangle=E_n|u_n\rangle$,
where $E_n\in [-\omega/2,\omega/2]$ is the $n$th quasienergy, and $|u_n\rangle$ is the corresponding Floquet state.
The time-evolution operator $\hat{U}_T$ over a Floquet period can characterize the stroboscopic dynamics,
$\psi(nT)={\hat{U}_{T}}^n\psi(0)$, at the stroboscopic times $nT$ ($n=1,2,...$).

Under this transformation $\hat{U}=\exp({i\sum_l lBt\hat{n}_l})$, one can demonstrate the forms of magnon density operator and two-magnon correlation operator are invariant.
Since the magnon density operator commutes with this unitary transformation, it naturally exists $\hat{U}\hat{n}_l\hat{U}^{{\dag}}=\hat{n}_l$.
Similarly, the two-magnon correlation operator turns to be $\hat{U}\hat{a}^{\dag}_{x}\hat{a}^{\dag}_{y} \hat{a}_{y}\hat{a}_{x}\hat{U}^{{\dag}}$ after the unitary transformation.
Due to $\hat{U}^{{\dag}}\hat{U}=\textbf{1}$, it allows us to obtain $\hat{U}\hat{a}^{\dag}_{x}\hat{a}^{\dag}_{y} \hat{a}_{y}\hat{a}_{x}\hat{U}^{{\dag}}=\hat{U}\hat{a}^{\dag}_{x}\hat{U}^{{\dag}}\hat{U}\hat{a}^{\dag}_{y} \hat{U}^{{\dag}}\hat{U} \hat{a}_{y}\hat{U}^{{\dag}}\hat{U}\hat{a}_{x}\hat{U}^{{\dag}}$.
According to the formula $e^{\hat{A}}\hat{B}e^{-\hat{A}}=\sum\limits_{n=0}^{\infty}\frac{1}{n!}[\hat{A}^{(n)},\hat{B}]$, it is easy to conclude that
\begin{eqnarray} \nn
\hat{U}\hat{a}^{\dag}_{x}\hat{U}^{{\dag}}&=&\exp({i\sum_l lBt\hat{n}_l})\hat{a}^{\dag}_{x}\exp(-{i\sum_l lBt\hat{n}_l})\\
&=&e^{ixBt}\hat{a}^{\dag}_{x}
\end{eqnarray}
and
\begin{eqnarray} \nn
\hat{U}\hat{a}_{x}\hat{U}^{{\dag}}&=&\exp({i\sum_l lBt\hat{n}_l})\hat{a}_{x}\exp(-{i\sum_l lBt\hat{n}_l})\\
&=&e^{-ixBt}\hat{a}_{x}.
\end{eqnarray}
Finally, we have
\begin{eqnarray} \nn
&&\hat{U}\hat{a}^{\dag}_{x}\hat{a}^{\dag}_{y} \hat{a}_{y}\hat{a}_{x}\hat{U}^{{\dag}}\\ \nn
&&=\exp({i\sum_l lBt\hat{n}_l})\hat{a}^{\dag}_{x}\hat{a}^{\dag}_{y} \hat{a}_{y}\hat{a}_{x}\exp(-{i\sum_l lBt\hat{n}_l})\\ \nn
&&=e^{ixBt}\hat{a}^{\dag}_{x}e^{iyBt}\hat{a}^{\dag}_{y}e^{-iyBt} \hat{a}_{y}e^{-ixBt}\hat{a}_{x}\\ \nn
&&=\hat{a}^{\dag}_{x}\hat{a}^{\dag}_{y} \hat{a}_{y}\hat{a}_{x}.\\
\end{eqnarray}

\subsection{Edge localizations in noninteracting systems \label{Sec31}}

\begin{figure}[htp]
\begin{center}
\includegraphics[width=0.45\textwidth]{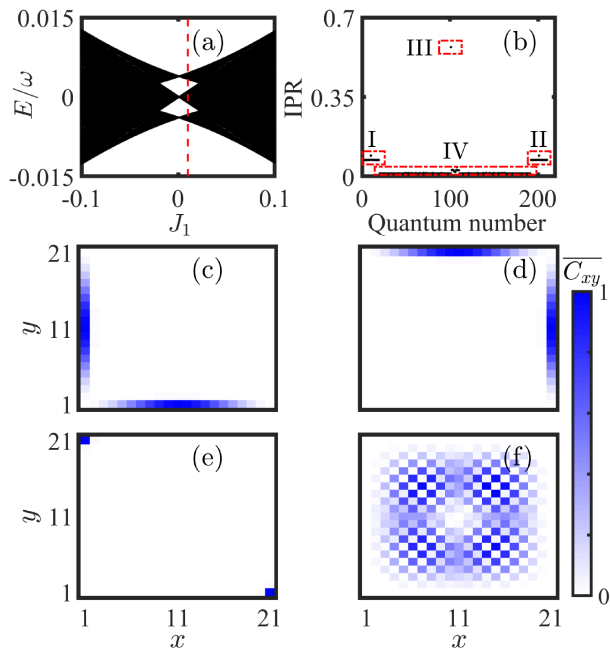}
\end{center}
\caption{(Color online)
(a) Quasienergy spectrum as a function of $J_1$ in the absence of interaction for parameters $\omega=B=8$ and $L=21$.
(b) The inverse participation ratio spectrum of the Floquet states corresponds to $J_1=0.01$ marked with red dashed line in (a).
The mode number on the horizontal axis is ordered for increasing values of the quasienergy.
(c)-(f) The normalized two-magnon correlations $\overline{C_{xy}}=C_{xy}/C^{max}_{xy}$ of the corresponding Floquet states respectively belonging to $(\rm{\uppercase\expandafter{\romannumeral1}})$, $(\rm{\uppercase\expandafter{\romannumeral2}})$, $(\rm{\uppercase\expandafter{\romannumeral3}})$ and $(\rm{\uppercase\expandafter{\romannumeral4}})$ in (b).
The energy unit is set as $J_0=\hbar=1$ and the coordinate unit is set as the lattice constant.}
\label{fig:noninteracting}
\end{figure}

Before proceeding to study the interplay between the two types of edge defects, we first address properties of two noninteracting magnons, as a natural extension of single-particle defect states~\cite{BZhu2020}.
Since there is no interaction, the interaction-induced edge defects also vanish and the two-excitation states can be approximately viewed as the combinations of two independent single-excited states, and their energies are approximately equal to the sum of energies of individual excitations.
Thus, prerequisite knowledge of single-particle states helps us understand the two-excitation states.
There are three kinds of single-particle states~\cite{BZhu2020}, extended state with energy $\varepsilon_k$, defect state at the left edge with energy $\varepsilon_{-}$ and defect states at the right edge with energy $\varepsilon_{+}=-\varepsilon_{-}$. We assume that the relative magnitudes of these energies satisfy $\varepsilon_{-}<\varepsilon_{k}<\varepsilon_{+}$.
Considering that double occupation is forbidden, there are four different combinations accordingly, (I) one is left-edge state and the other is extended state with total energy around $\varepsilon_{k}+\varepsilon_{-}\in [-J_1/2+\varepsilon_{-},J_1/2+\varepsilon_{-}]$, (II) one is right-edge state and the other is extended state with total energy around $\varepsilon_{k}+\varepsilon_{+}\in [-J_1/2+\varepsilon_{+},J_1/2+\varepsilon_{+}]$, (III) both are edge states with total energy  $\varepsilon_{-}+\varepsilon_{+}=0$, and (IV) both are extended states with total energy around $\varepsilon_{k}+\varepsilon_{k'}\in [-J_1,J_1]$.
Hence the type-(III) state penetrates into the continuum of type-(IV) states, but the relative energy magnitudes of the other two-excitation states depend on the parameters.
In Fig.~\ref{fig:noninteracting}(a), we show the quasienergy spectrum as a function of $J_1$ obtained from the Hamiltonian~\eqref{staticHamitonian}.
The other parameters are chosen as $\Delta=0$, $\omega=B=8$ and $L=21$.
It is obvious that $\varepsilon_{k}+\varepsilon_{-}<\varepsilon_{k}+\varepsilon_{k'}\approx \varepsilon_{-}+\varepsilon_{+}<\varepsilon_{k}+\varepsilon_{+}$ when $J_1$ approaches to zero.
However, these bands overlap with each other when $|J_1|$ is large enough.
Regarding the Floquet-Wannier-Zeeman spin chain~\eqref{originalspinchain}, the edge localization with $J_1=0$ in Fig.~\ref{fig:noninteracting}(a) essentially comes from the Wannier-Zeeman localization~\cite{VVGann367222010,YAKosevich252460022013,YKe910534092015}.
While for $J_1\neq0$, the edge localization in Fig.~\ref{fig:noninteracting}(a) is the joint effect of the gradient magnetic field and periodically-modulated spin-exchange strength.
It means that the Floquet-Wannier-Zeeman edge states change to Wannier-Zeeman edge states once $J_1=0$.

To separate these four types of states, we compute the inverse participation ratios (IPRs) for all of the Floquet states from the perspective of the localization properties.
For an arbitrary two-excitation Floquet state  $|u_n\rangle=\sum_{l_1<l_2}u_{l_1l_2}^n|l_1l_2\rangle$, the IPR~\cite{LWang2017} is defined as
\begin{equation}
{\rm IPR}^{(n)}=\frac{\sum_{l_1<l_2}|u^n_{l_1l_2}|^4}{\big(\sum_{l_1<l_2}|u^n_{l_1l_2}|^2\big)^2},
\end{equation}
which quantifies the localization degree. ${\rm IPR}\sim 1$ for the mostly localized states while ${\rm IPR}\sim 0$ for the most extended states.
For $J_1=0.01$, the IPRs of all Floquet states are shown in Fig.~\ref{fig:noninteracting}(b).
The IPR of type-(III) state is the largest,  IPRs of type-(I) and type-(II) states are the median, and IPRs of type-(IV) states are the smallest.
Though the quasienergy of type-(III) state is mixed with those of type-(IV) states, their difference in IPRs is gigantic.

We analyze the correlation properties of the four types of states under the guidance of the IPRs.
In Fig.~\ref{fig:noninteracting}(c), we calculate the two-magnon correlation of the $1$-th state belonging to type-$(\rm{\uppercase\expandafter{\romannumeral1}})$.
It reveals that a magnon tends to stay at the left endpoint of the spin chain while the other magnon freely distributes in the bulk lattice.
Similarly, the two-magnon correlation of the $210$-th state belonging to type-$(\rm{\uppercase\expandafter{\romannumeral2}})$ indicates that a magnon is localized at the right endpoint while another magnon freely distributes in the bulk lattice; see Fig.~\ref{fig:noninteracting}(d).
After extended calculations, we find that the correlation properties of the other type-$(\rm{\uppercase\expandafter{\romannumeral1}})$ and type-$(\rm{\uppercase\expandafter{\romannumeral2}})$  states are similar to those in Figs.~\ref{fig:noninteracting}(c) and~\ref{fig:noninteracting}(d), respectively.
These states are classified as one-magnon edge states.
Type-$(\rm{\uppercase\expandafter{\romannumeral3}})$ state with a maximum IPR in the continuum is presented in Fig.~\ref{fig:noninteracting}(e).
As expected, the correlation shows that such state is mostly distributing on two outermost sites, the so-called two-magnon EIC.
The correlation of a type-(IV) state is shown in Fig.~\ref{fig:noninteracting}(f), where both of the two magnons freely distribute in the bulk lattice.

\subsection{Effective model for single-magnon dynamics \label{Sec32}}

\begin{figure}[htp]
\begin{center}
\includegraphics[width=0.45\textwidth]{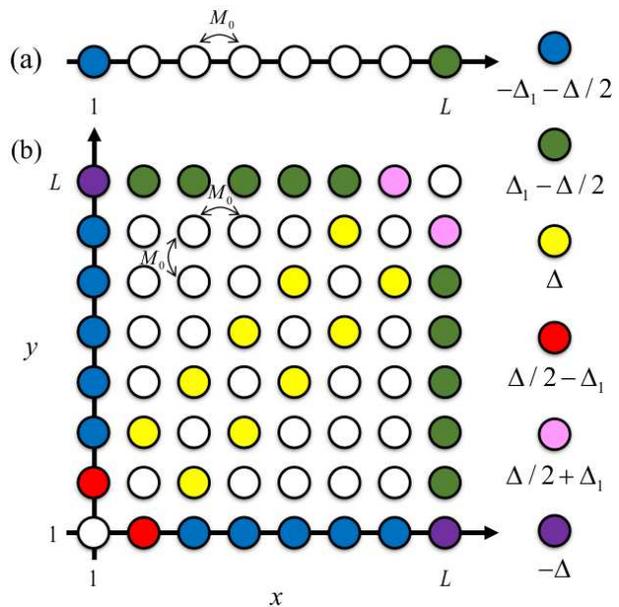}
\end{center}
\caption{(Color online)
Schematic representations of static effective models based on the multiscale perturbation analysis.
(a) Sketch of the static effective model~\eqref{Onemultitimeequation}.
It amounts to load single magnon into a lattice with a renormalized tunneling rate and effective defects at two endpoints.
(b) The 2D schematic diagram of the static effective model~\eqref{Twomultitimeequation} based on the mapping of a 1D two-magnon system onto a 2D single-magnon system.
The coordinate unit is set as the lattice constant.}
\label{fig:staticeffectivemodel}
\end{figure}

For $\Delta\neq0$, we analytically derive a static effective model with the multiscale perturbation analysis~\cite{ILGaranovich2008,AlexSzameit2008,SLonghi2008,SLonghi2013,GDellaVallePRB2014,BZhu2020} for single-magnon systems.
Multiscale perturbation analysis has successfully explained the effective edge defects in the curved waveguide~\cite{ILGaranovich2008} which have been verified in experiments~\cite{AlexSzameit2008}.
In the high-frequency limit, the modulation period $T$ is small and so that one can introduce a small parameter $\varepsilon$ with $T=O(\varepsilon)$.
Under the condition of the small parameter $\varepsilon$, the time $t$ can be expanded as a multiple time scale with $t=\varepsilon^{-1}t_{-1}+\varepsilon^{0}t_{0}+\varepsilon t_{1}+\varepsilon^{2}t_{2}+\cdot\cdot\cdot$.
The series expansion of the time is regarded as multiscale.
After substituting the multiple time scales into the time-evolution equation, the effective model can be obtained by collecting different orders of $\varepsilon$.
Our static effective model is derived up to first-order time scale based on the multiscale perturbation analysis.

A general wave-function for the case of single magnon is given by $|\psi\rangle=\sum_{l}\psi_{l}|l\rangle$ with the probability amplitude $\psi_{l}=\langle\textbf{0}|\hat{a}_{l}|\psi\rangle$.
After substituting the single-magnon state into the Schr\"{o}dinger equation
\begin{equation}
i\frac{d}{dt}|\psi(t)\rangle=\hat{\mathcal{H}}^{rot}_m|\psi(t)\rangle,
\end{equation}
the time evolution of probability amplitudes in the single-magnon basis yields
\begin{equation} \label{equation14}
i \frac{d \psi_{l}(t)}{d t}=\sum_{m} W(t;l,m) \psi_{m}(t)
\end{equation}
where
\begin{equation}
\begin{aligned}
W(t;l,m)=&\left(1-\delta_{l,1}\right) \delta_{m, l-1} \mathcal{J}(t)+\left(1-\delta_{l, L}\right) \delta_{m, l+1} \mathcal{J}^{*}(t) \\
&-\frac{\Delta}{2}\delta_{l,1}\delta_{l, m}-\frac{\Delta}{2}\delta_{l,L}\delta_{l, m}
\end{aligned}
\end{equation}
for the open boundary condition.
$\delta_{l,m}$ is the Kronecker delta function.
Due to the time periodicity, we have $W(t ; l, m)=W(t+T ; l, m)$.
In the high-frequency case, $\omega\gg J_0,J_1$, one can introduce a small parameter $\varepsilon$, corresponding a small modulation period $T=O(\varepsilon)$.
The time $t$ can be expanded as the multiple time scales $t=\varepsilon^{-1}t_{-1}+\varepsilon^{0}t_{0}+\varepsilon t_{1}+\varepsilon^{2}t_{2}+\cdot\cdot\cdot$, according to the small parameter $\varepsilon$.
Then we utilize a series expansion
\begin{equation} \label{seriesexpansion}
\begin{aligned}
\psi_{l}(t)=& U_{l}\left(t_{0}, t_{1}, t_{2}, \ldots\right)+\varepsilon v_{l}\left(t_{-1}, t_{0}, t_{1}, t_{2}, \ldots\right) \\
&+\varepsilon^{2} w_{l}\left(t_{-1}, t_{0}, t_{1}, t_{2}, \ldots\right) \\
&+\varepsilon^{3} \zeta_{l}\left(t_{-1}, t_{0}, t_{1}, t_{2}, \ldots\right)+O\left(\varepsilon^{4}\right)
\end{aligned}
\end{equation}
as the solution of Eq.~\eqref{equation14}, where $t_{n}=\varepsilon^{n} t$.
Under the condition of the series expansion, the time differential turns to be
\begin{equation}
\frac{d}{d t}=\varepsilon^{-1} \frac{\partial}{\partial t_{-1}}+\varepsilon^{0}\frac{\partial}{\partial t_{0}}+\varepsilon^{1} \frac{\partial}{\partial t_{1}}+\varepsilon^{2} \frac{\partial}{\partial t_{2}}+\cdots.
\end{equation}
After introducing a notation
\begin{equation}\label{notation}
\langle\bullet\rangle=\varepsilon T^{-1} \int_{\varepsilon^{-1} t}^{\varepsilon^{-1}(t+T)}(\bullet)\left(t_{-1}\right) d t_{-1},
\end{equation}
$U_l$ describes the averaged behavior over a modulation period as
\begin{equation} \label{equation19}
\left\langle\psi_{l}\right\rangle=U_{l}, \quad\left\langle\frac{d \psi_{l}}{d t}\right\rangle=\frac{d U_{l}}{d t}.
\end{equation}
Since $U_l$ is independent of the fast variable $t_{-1}$, the slowly varying component $U_l$ satisfies
\begin{equation}\label{equation20}
\left\langle U_{l}\right\rangle=U_{l}, \quad\left\langle\frac{d U_{l}}{d t}\right\rangle=\frac{d U_{l}}{d t}.
\end{equation}
Combining Eqs.~\eqref{equation19} and~\eqref{equation20}, one can obtain
\begin{eqnarray}
\left\langle v_{l}\right\rangle&=&\left\langle w_{l}\right\rangle=\left\langle\zeta_{l}\right\rangle \equiv 0, \nonumber \\ \quad\left\langle\frac{\partial v_{l}}{\partial t_{n}}\right\rangle&=&\left\langle\frac{\partial w_{l}}{\partial t_{n}}\right\rangle=\left\langle\frac{\partial \zeta_{l}}{\partial t_{n}}\right\rangle \equiv 0,
\end{eqnarray}
for $n=-1,0,1,2,...$.

We substitute Eq.~\eqref{seriesexpansion} into Eq.~\eqref{equation14} and collect terms up to the first order of $\varepsilon$.
In the zero-order term $\varepsilon^{0}$, we have
\begin{equation}
\begin{aligned}
W_{0}(l, m) &=\langle W(t ; l, m)\rangle \\
&=\left(1-\delta_{l,1}\right) \delta_{m, l-1} M_0+\left(1-\delta_{l, L}\right) \delta_{m, l+1}M_0\\
&-\frac{\Delta}{2}\delta_{l, 1}\delta_{l, m}-\frac{\Delta}{2}\delta_{l, L}\delta_{l, m}.
\end{aligned}
\end{equation}
It clearly shows that the zero-order term amounts to make a rotating-wave approximation to the Floquet magnon Hamiltonian~\eqref{FloquetH} in the high-frequency limit.
The first order $\varepsilon^{1}$ reads
\begin{equation}
\begin{aligned}
\sum_{j} W_{1}(l, j, m) &=i \sum_{j}\langle W(t ; l, j) M(t ; j, m)\rangle \\
&=-\delta_{l,1} \Delta_1\delta_{l,m}+\delta_{l, L} \Delta_1\delta_{l,m}
\end{aligned}
\end{equation}
with $M(t ; j, m)=\int_{0}^{t}\left[W\left(t^{\prime} ; j, m\right)-W_{0}(l, m)\right] d t^{\prime}$ and $\Delta_1=\big(|M_1|^2+|M_2|^2/2\big)/\omega$.
The first-order term brings about the effective on-site potentials at the two endpoints with the same values and opposite signs.
In the high-frequency conditions, the higher-order terms play an unimportant role which could be neglected.
Using the notation~\eqref{notation}, the time evolution of $U_l$ is given by
\begin{equation}
i \frac{d U_{l}}{d t}=\sum_{m} W_{s}(l, m) U_{m}
\end{equation}
where the effective couplings are
\begin{equation}
W_{s}(l, m)=W_{0}(l, m)+\sum_{j} W_{1}(l, j, m).
\end{equation}
Finally, by implementing the multiscale perturbation analysis in the high-frequency regime $\omega\gg J_0, J_1$, we obtain a static effective model up to
the first-order time scale
\begin{eqnarray}\label{Onemultitimeequation} \nn
i\frac{d}{d t}U_{l}&=&M_0(U_{l-1}+U_{l+1})\\
&-&(\frac{\Delta}{2}+\Delta_1)\delta_{l,1}U_{l}-(\frac{\Delta}{2}-\Delta_1)\delta_{l,L}U_{l}.
\end{eqnarray}
We find that the effective tunneling strength $M_0$ depends on the modulation amplitude $J_1$.
In addition to the interaction-induced edge defects $-\Delta/2$, there is another kind of edge defects $|\Delta_1|$ with opposite values at the outermost sites.
Since such edge defects $|\Delta_1|$ originate from the joint effect of the gradient magnetic field and periodically-modulated spin-exchange strength in the Floquet-Wannier-Zeeman spin chain~\eqref{originalspinchain}, we call them as Floquet-Wannier-Zeeman edge defects.
The interplay between interaction-induced edge defects and Floquet-Wannier-Zeeman edge defects attributes to the joint effects among interaction, gradient magnetic field and periodically-modulated spin-exchange strength.
Modulation frequency and modulation amplitude provide more possibilities to tune the edge defects.
It is worth emphasizing that when $M_0=M_2=0$ for $J_1=0$, the Floquet-Wannier-Zeeman edge defects change to Wannier-Zeeman edge defects $\Delta_1=|M_1|^2/B=J_0^2/(4B)$ but the effective tunneling also vanishes.
The Wannier-Zeeman edge defects via the multiscale perturbation analysis agree with the extra energy shifts at the endpoints in a Wannier-Zeeman system (see Appendix~\ref{appendixA} for more details).

The static effective model~\eqref{Onemultitimeequation} is schematically presented in Fig.~\ref{fig:staticeffectivemodel}(a), which describes a magnon moves on a lattice with a renormalized tunneling strength and two different types of defects at the endpoints.
The static effective model~\eqref{Onemultitimeequation} indicates that, a competition between the interaction-induced edge defects and Floquet-Wannier-Zeeman edge defects makes it possible to manipulate the magnon transport and tune interaction-induced bound-magnon edge state into the scattering-state band.

\begin{figure}[htp]
\begin{center}
\includegraphics[width=0.45\textwidth]{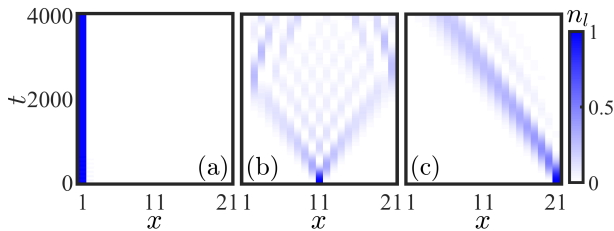}
\end{center}
\caption{(Color online)
The time evolution of magnon distribution obtained from the Hamiltonian~\eqref{staticHamitonian}.
The magnon is initially prepared at left endpoint (a), center (b) and right endpoint (c) for $\Delta=2\Delta_1$, $J_1=0.01$ and $\omega=B=8$.
The length of chain is $L=21$.
The coordinate unit is set as the lattice constant and the time unit is set as $2\pi/J_0$.}
\label{fig:asymmtricevolution}
\end{figure}

For $\Delta=2\Delta_1$, a single magnon initially prepared in the left endpoint will stay localized in the time evolution, indicating that the left endpoint has a strong defect potential; see Fig.~\ref{fig:asymmtricevolution}(a).
However, delocalized transport appears when loading a magnon in the right endpoint; see Fig.~\ref{fig:asymmtricevolution}(c).
This is because the two types of edge defects cancel with each other, that is, in the condition $\Delta=2\Delta_1$ the right endpoint has the same onsite energy as the bulk lattice and hence can not trap the excitation any more.
When loading a magnon in a bulk site, the magnon will undergo quantum walks; see Fig.~\ref{fig:asymmtricevolution}(b).
This is because the modulated frequency is resonant with the tilted potential and resonant tunneling happens~\cite{WLiu1911,VVIvanov2008,MGTarallo2012}.
In addition, a significant asymmetric transport is observed in Fig.~\ref{fig:asymmtricevolution}(b), that a repulsion effect occurs at the left boundary while a collision effect at the right boundary for a magnon setting out from a bulk site.

\subsection{Effective model for bound-magnon edge states in the continuum \label{Sec33}}

Analogously, we perform the multiscale perturbation analysis for two interacting magnons when $\omega\gg J_0, J_1$.
The slowing varying components $U_{l_1l_2}$ of two-body wave functions up to
the first order satisfy
\begin{eqnarray}\label{Twomultitimeequation} \nn
i\frac{d}{d t}U_{l_1l_2}&=&M_0(U_{l_1,l_2-1}+U_{l_1-1,l_2}+U_{l_1,l_2+1}+U_{l_1+1,l_2})\\ \nn
&+&\Delta\delta_{l_1,l_2\pm 1}U_{l_1l_2}-(\frac{\Delta}{2}+\Delta_1)\delta_{l_1,1}U_{l_1l_2}\\
&-&(\frac{\Delta}{2}-\Delta_1)\delta_{l_2,L}U_{l_1l_2}.
\end{eqnarray}
This model can be mapped to a magnon hopping in a two-dimensional square lattice, where the two-dimensional positions correspond to the positions of two magnons~\cite{LinHuLi2020,ChingHuaLee2020}, as sketched in Fig.~\ref{fig:staticeffectivemodel}(b).
In order to present each terms of static effective model~\eqref{Twomultitimeequation} more clearly, only $8\times8$ sites are chosen.
The different onsite energies are denoted by different colors, which are more complex than those in Fig.~\ref{fig:staticeffectivemodel}(a).
Two-site nearest-neighbor interaction $\Delta$ in the 1D system is translated into a local potential $\Delta$ on the minor-diagonal lines $x=y\pm1$.

\begin{figure}[htp]
\begin{center}
\includegraphics[width=0.45\textwidth]{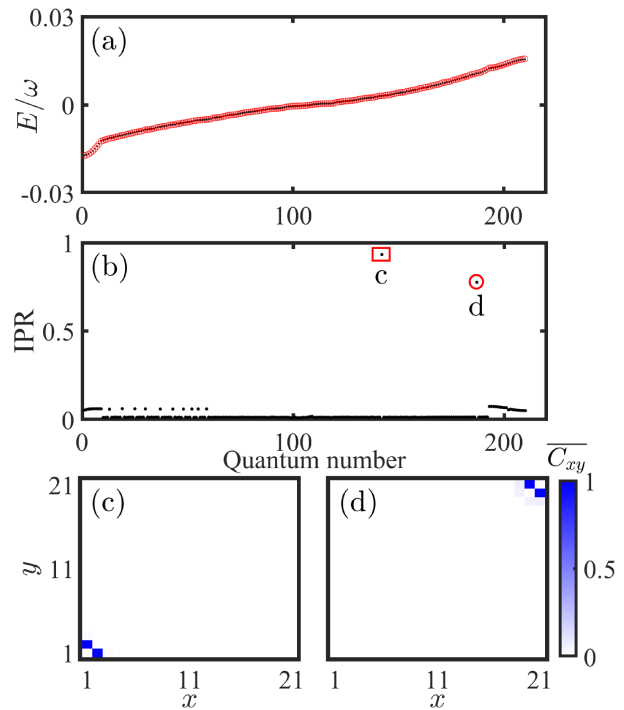}
\end{center}
\caption{(Color online)
(a) Quasienergy spectrum in ascending order for the values of the quasienergies.
The black dots denote the quasienergies for the Hamiltonian~\eqref{staticHamitonian} and the red circles are the eigenenergies for the static effective model~\eqref{Twomultitimeequation}.
The parameters are chosen as $\Delta=0.1$, $J_1=0.1$, $\omega=B=8$ and $L=21$.
(b) The corresponding inverse participation ratio spectrum of the Floquet states for the Hamiltonian~\eqref{staticHamitonian} marked with black dots in (a). The mode number on the horizontal axis is ordered for increasing values of the quasienergies.
The normalized two-magnon correlations $\overline{C_{xy}}=C_{xy}/C^{max}_{xy}$ of the corresponding BEICs identified in (b) are clearly visible on the endpoints of minor-diagonal lines $x=y\pm1$ in (c) and (d).
The coordinate unit is set as the lattice constant.}
\label{fig:BICS}
\end{figure}

In the subsection~\ref{Sec22} we find bound-magnon edge state due to the interaction.
Here, we try to understand how the coexistence of interaction, gradient magnetic field and periodically-modulated spin-exchange strength affects these bound-magnon edge states.
Because the Floquet-Wannier-Zeeman edge defects are weak in general, we consider the weak interaction and calculate the quasienergies in an ascending order of their values; see Fig.~\ref{fig:BICS}(a).
The parameters are chosen as $\Delta=0.1$, $J_1=0.1$, $\omega=B=8$ and $L=21$.
The black dots and red circles are respectively obtained by diagonalizing the Hamiltonian~\eqref{staticHamitonian} and the static effective model~\eqref{Twomultitimeequation}, which are well consistent with each other.
It means that the static effective model~\eqref{Twomultitimeequation} can well capture the properties of two-magnon Floquet states.
Since the bound-magnon edge states are highly localized with large IPRs, we calculate the IPRs of all the Floquet states from the Hamiltonian~\eqref{staticHamitonian}, as shown in Fig.~\ref{fig:BICS}(b).
We find two bound-magnon edge states with large IPRs do appear marked with c and d in Fig.~\ref{fig:BICS}(b), characterized by a predominant correlated distribution around endpoint of the minor-diagonal lines ($x=y\pm1$); see Figs.~\ref{fig:BICS}(c) and~\ref{fig:BICS}(d), respectively.
We name these states as BEICs, because they have energies around $\Delta/2\pm \Delta_1$ mixed with the continuum band of scattering states which ranges from $-J_1$ to $J_1$.
Compared to the degenerate bound-magnon edge states in Fig.~\ref{fig:twomagnonspectrum}(c), we can clearly find that the energies of BEICs are non-degenerate and shifted.
One may also tune one bound-magnon edge state to be BEIC and the other out of continuum as well as both the bound-magnon edge states out of continuum by changing $J_1$.
Thus, Floquet-Wannier-Zeeman manipulation provides a powerful tool to engineer these defect states.

\section{Summary \label{Sec4}}

We investigate the edge states of magnon excitations in Floquet-Wannier-Zeeman spin chains under open boundary conditions.
Due to two-site nearest-neighbor interaction, there appear edge defects which can trap magnons at the edges and repulse magnons from the bulk.
We propose to distinguish different isolated bands of the energy spectrum by correlation functions.
The Floquet-Wannier-Zeeman edge defects are created by introducing a gradient magnetic field and a periodically-modulated spin exchange strength in a resonantly driven condition.
We explore the interplay between interaction-induced edge defects and Floquet-Wannier-Zeeman edge defects based on the multiscale perturbation analysis.
More specifically, relying on the choice of parameters, the interaction-induced bound-magnon edge state can be manipulated into bound-magnon edge state in the continuum and a symmetric magnon transport is tuned to be asymmetric.
Our work adds an important piece to the jigsaw puzzle of the quantum spin chain.
It would also be interesting to extend our results to topological systems for exploring how the interaction-induced defects affect the topological edge states.


\begin{acknowledgments}
W. Liu and B. Zhu made equal contributions.
The authors thank Xizhou Qin and Ling Lin for discussions.
This work is supported by the NSFC (Grants No. 12025509, No. 11874434), the Key-Area Research and Development Program of GuangDong Province (Grants No. 2019B030330001), and the Science and Technology Program of Guangzhou (China) (Grants No. 201904020024).
Y.K. is partially supported by the Office of China Postdoctoral Council (Grant No. 20180052) and the National Natural Science Foundation of China (Grant No. 11904419).
\end{acknowledgments}
\appendix
\section{Wannier-Zeeman edge defects} \label{appendixA}

Interestingly, even though $M_0=M_2=0$ for $J_1=0$, there still exists defects with strength $\Delta_1=|M_1|^2/B=J_0^2/(4B)$ which intrinsically belong to Wannier-Zeeman edge defects.
To better understand it, we turn to analyze the Wannier-Zeeman system of a single magnon in the titled model without interaction and periodically-modulated spin-exchange strength
\begin{eqnarray}\label{singlemagnontitledmodel}
\hat{\mathcal{H}}^{a}_m=\sum\limits_{l=1}^{L-1} \left(\frac{J_0}{2}\hat{a}^{\dag}_l\hat{a}_{l+1}+{\rm H.c.}\right)+\sum\limits_{l=1}^{L} Bl\hat{n}_l.
\end{eqnarray}
For $B\gg J_0$, the tilted potential $\hat{\mathcal{H}}^{a}_0=\sum_{l} Bl\hat{n}_l$ serves as a dominant term while the tunneling $\hat{\mathcal{H}}^{a}_1=\sum_{l} (\frac{J_0}{2}\hat{a}^{\dag}_l\hat{a}_{l+1}+{\rm H.c.})$ is a perturbation.
The $L$ eigenstates $\{|l\rangle\}$ of $\hat{\mathcal{H}}^{a}_0$ are divided into bulk eigenstates with $l\neq1,L$ and edge eigenstates with $l=1,L$, and the eigenvalues of $\hat{\mathcal{H}}^{a}_0$ are $E_l=lB$.
For an edge eigenstate with $l=1$, its corresponding projection operator is
\begin{equation}
\hat{P}=|l\rangle\langle l|.
\end{equation}
The projection operator of the remaining $L-1$ eigenstates can be written as
\begin{equation}
\hat{S}=\sum_{k \neq l} \frac{1}{l B-kB}|k\rangle\langle k|.
\end{equation}
According to the second-order perturbation theory~\cite{SBravyi32627932011,MTakahashi1012891977}, the effective model is given by
\begin{equation}
\hat{\mathcal{H}}^{a}_{m,\text {eff }}=\hat{h}_{0}+\hat{h}_{1}+\hat{h}_{2}=E_{l} \hat{P}+\hat{P} \hat{\mathcal{H}}^{a}_1 \hat{P}+\hat{P} \hat{\mathcal{H}}^{a}_1 \hat{S}\hat{\mathcal{H}}^{a}_1 \hat{P}.
\end{equation}
For the lowest order and first order, we have
\begin{equation}
\hat{h}_{0}=E_{l} \hat{P}=lB |l\rangle\langle l|,
\end{equation}
and
\begin{equation}
\hat{h}_{1}=\hat{P} \hat{\mathcal{H}}_{1}^{a} \hat{P}=0.
\end{equation}
The second-order term satisfies
\begin{equation}
\hat{h}_{2}=\hat{P} \hat{\mathcal{H}}_{1}^{a} \hat{S} \hat{\mathcal{H}}_{1}^{a} \hat{P}=-\frac{J_0^2}{4B}|l\rangle\langle l|.
\end{equation}
Therefore, the effective model for state $|1\rangle$ up to second order is given by
\begin{equation}\label{leftenergyshift}
\hat{\mathcal{H}}_{m, \mathrm{eff}}^{a}=(B-\frac{J_0^2}{4B})|1\rangle\langle 1|.
\end{equation}
Further, the second-order effective model for state $|L\rangle$ yields
\begin{equation}\label{rightenergyshift}
\hat{\mathcal{H}}_{m, \mathrm{eff}}^{a}=(LB+\frac{J_0^2}{4B})|L\rangle\langle L|.
\end{equation}
Similarly, we also perform the second-order perturbation analysis to the state $|l\rangle$ with $l\neq1,L$, and the corresponding effective model turns to be
\begin{equation}
\hat{\mathcal{H}}_{m, \mathrm{eff}}^{a}=l B|l\rangle\langle l|.
\end{equation}
The extra energy $-J_0^2/(4B)$ [$J_0^2/(4B)$] at state $|1\rangle$ ($|L\rangle$) comes from a second-order coupling between state $|1\rangle$ ($|L\rangle$) and state $|2\rangle$ ($|L-1\rangle$).
While for a state $|l\rangle$ with $l\neq1,L$, the extra energy of the second-order coupling to state $|l+1\rangle$ cancels with the one to state $|l-1\rangle$.
The above effective models clearly reveal that there exist an edge defect $-J_{0}^{2}/(4 B)$ at the left endpoint and an edge defect $J_{0}^{2}/(4 B)$ at the right endpoint, which agree with the Wannier-Zeeman edge defect $\Delta_{1}=J_{0}^{2} /(4 B)$ of multiscale perturbation analysis for $J_1=0$.
In addition, our results Eqs.~\eqref{leftenergyshift} and~\eqref{rightenergyshift} also agree with the extra energy shifts at the boundaries in a finite tilted chain~\cite{JHeinrichs521491972} when the hopping strength is far less than the tilted field.

\bibliographystyle{apsrev4-1}

\begin{thebibliography}{99}
%
\bibitem{USchollwock2004}
U. Schollw\"ock, J. Richter, D. J. J. Farnell, R. F. Bishop (Eds.), Quantum Magnetism, Lect. Notes Phys. \textbf{645} (Springer, Berlin, Heidelberg, 2004).
%
\bibitem{SSachdev1999}
S. Sachdev, Quantum Phase Transitions (Cambridge University Press, Cambridge, England, 1999).
%
\bibitem{DWRobinson1967}
D. W. Robinson. The statistical mechanics of quantum spin systems, Comm. Math. Phys. \textbf{6}, 151 (1967).
%
\bibitem{MAnderlini2007}
M. Anderlini, P. J. Lee, B. L. Brown, J. Sebby-Strabley, W. D. Phillips, and J. V. Porto, Controlled exchange interaction between pairs of neutral atoms in an optical lattice, Nature (London) \textbf{448}, 452 (2007).
%
\bibitem{JSimon2011}
J. Simon, W. S. Bakr, R. Ma, M. E. Tai, P. M. Preiss, and M. Greiner, Quantum simulation of antiferromagnetic spin chains in an optical lattice, Nature (London) \textbf{472}, 307 (2011).
%
\bibitem{JStruck2011}
J. Struck, C. \"{O}lschl\"{a}ger, R. Le Targat, P. Soltan-Panahi, A. Eckardt, M. Lewenstein, P. Windpassinger, and K. Sengstock, Quantum Simulation of Frustrated Classical Magnetism in Triangular Optical Lattices, Science \textbf{333}, 996 (2011).
%
\bibitem{TFukuhara20131}
T. Fukuhara, A. Kantian, M. Endres, M. Cheneau, P. Schau{\ss}, S. Hild, D. Bellem, U. Schollw\"{o}ck, T. Giamarchi, C. Gross,
I. Bloch, and S. Kuhr, Quantum dynamics of a mobile spin impurity, Nat. Phys. \textbf{9}, 235 (2013).
%
\bibitem{TFukuhara20132}
T. Fukuhara, P. Schau{\ss}, M. Endres, S. Hild, M. Cheneau, I. Bloch, and C. Gross, Microscopic observation of magnon bound
states and their dynamics, Nature (London) \textbf{502}, 76 (2013).
%
\bibitem{AFriedenauer2008}
A. Friedenauer, H. Schmitz, J. T. Glueckert, D. Porras, and T. Schaetz, Simulating a quantum magnet with trapped ions, Nat. Phys. \textbf{4}, 757 (2008).
%
\bibitem{KKim2009}
K. Kim, M.-S. Chang, R. Islam, S. Korenblit, L.-M. Duan, and C. Monroe, Entanglement and Tunable Spin-Spin Couplings between Trapped Ions Using Multiple Transverse Modes, Phys. Rev. Lett. \textbf{103}, 120502 (2009).
%
\bibitem{EEEdwards2010}
E. E. Edwards, S. Korenblit, K. Kim, R. Islam, M.-S. Chang, J. K. Freericks, G.-D. Lin, L.-M. Duan, and C. Monroe, Quantum simulation and phase diagram of the transverse-field Ising model with three atomic spins, Phys. Rev. B \textbf{82}, 060412(R) (2010).
%
\bibitem{KKim2011}
K. Kim, S. Korenblit, R. Islam, E. E. Edwards, M.-S. Chang, C. Noh, H. Carmichael, G.-D. Lin, L.-M. Duan, C. C. J. Wang, J. K. Freericks, and C. Monroe, Quantum simulation of the transverse Ising model with trapped ions, New J. Phys. \textbf{13}, 105003 (2011).
%
\bibitem{JQYou2010}
J. Q. You, X.-F. Shi, X. Hu, and F. Nori, Quantum emulation of a spin system with topologically protected ground states using superconducting quantum circuits, Phys. Rev. B \textbf{81}, 014505 (2010).
%
\bibitem{ULasHeras2014}
U. Las Heras, A. Mezzacapo, L. Lamata, S. Filipp, A. Wallraff, and E. Solano, Digital Quantum Simulation of Spin Systems in Superconducting Circuits, Phys. Rev. Lett. \textbf{112}, 200501 (2014).
%
\bibitem{YSalathe2015}
Y. Salath\'{e}, M. Mondal, M. Oppliger, J. Heinsoo, P. Kurpiers, A. Poto\v{c}nik, A. Mezzacapo, U. Las Heras, L. Lamata, E. Solano, S. Filipp, and A. Wallraff, Digital Quantum Simulation of Spin Models with Circuit Quantum Electrodynamics, Phys. Rev. X \textbf{5}, 021027 (2015).
%
\bibitem{RBarends2016}
R. Barends, A. Shabani, L. Lamata, J. Kelly, A. Mezzacapo, U. L. Heras, R. Babbush, A. G. Fowler, B. Campbell, Y. Chen, Z. Chen, B. Chiaro,
A. Dunsworth, E. Jeffrey, E. Lucero, A. Megrant, J. Y. Mutus, M. Neeley, C. Neill, P. J. J. O'Malley, C. Quintana, P. Roushan, D. Sank, A.
Vainsencher, J. Wenner, T. C. White, E. Solano, H. Neven, and J. M. Martinis, Digitized adiabatic quantum computing with a
superconducting circuit, Nature \textbf{534}, 222 (2016).
%
\bibitem{GWendin2017}
G. Wendin, Quantum information processing with superconducting circuits: a review, Rep. Prog. Phys. \textbf{80}, 106001 (2017).
%
\bibitem{XPeng2005}
X. Peng, J. Du, and D. Suter, Quantum phase transition of ground-state entanglement in a Heisenberg spin chain simulated
in an NMR quantum computer, Phys. Rev. A \textbf{71}, 012307 (2005).
%
\bibitem{XPeng2010}
X. Peng, S. Wu, J. Li, D. Suter, and J. Du, Observation of the Ground-State Geometric Phase in a Heisenberg $XY$ Model, Phys. Rev. Lett. \textbf{105}, 240405 (2010).
%
\bibitem{ZLi2014}
Z. Li, H. Zhou, C. Ju, H. Chen, W. Zheng, D. Lu, X. Rong, C. Duan, X. Peng, and J. Du, Experimental Realization of a Compressed Quantum Simulation of a 32-Spin Ising Chain, Phys. Rev. Lett. \textbf{112}, 220501 (2014).
%
\bibitem{HBethe1931}
H. Bethe, Zur Theorie der Metalle, Z. Phys. \textbf{71}, 205 (1931).
%
\bibitem{MDate1966}
M. Date and M. Motokawa, Spin-Cluster Resonance in $\rm CoCl_{2}\cdot 2H_{2}O$, Phys. Rev. Lett. \textbf{16}, 1111 (1966).
%
\bibitem{JBTorrance1969}
J. B. Torrance and M. Tinkham, Excitation of multiple-magnon bound states in $\rm CoCl_{2}\cdot 2H_{2}O$, Phys. Rev. \textbf{187}, 595 (1969).
%
\bibitem{RHoogerbeets1984}
R. Hoogerbeets, A. J. van Duyneveldt, A. C. Phaff, C. H. W. Sw\"{u}ste, and W. J. M. de Jonge, Evidence for magnon bound-state excitations in the quantum chain system $(\rm C_{6}H_{11}NH_{3})CuCl_{3}$, J. Phys. C \textbf{17}, 2595 (1984).
%
\bibitem{IAffleck1987}
I. Affleck, T. Kennedy, E. H. Lieb, and H. Tasaki, Rigorous results on valence-bond ground states in antiferromagnets, Phys. Rev. Lett. \textbf{59}, 799 (1987).
%
\bibitem{FMei2018}
F. Mei, G. Chen, L. Tian, S.-L. Zhu, and S. Jia, Robust quantum state transfer via topological edge states in superconducting qubit chains, Phys. Rev. A \textbf{98}, 012331 (2018).
%
\bibitem{XLi2018}
X. Li, Y. Ma, J. Han, T. Chen, Y. Xu, W. Cai, H. Wang, Y. P. Song, Z.-Y. Xue, Z.-Q. Yin, and L. Sun, Perfect Quantum State Transfer in a Superconducting Qubit Chain with Parametrically Tunable Couplings, Phys. Rev. Applied \textbf{10}, 054009 (2018).
%
\bibitem{AZwick2011}
A. Zwick and O. Osenda, Quantum state transfer in a $XX$ chain with impurities, J. Phys. A \textbf{44}, 105302 (2011).
%
\bibitem{TJGApollaro2006}
T. J. G. Apollaro and F. Plastina, Entanglement localization by a single defect in a spin chain, Phys. Rev. A \textbf{74}, 062316 (2006).
%
\bibitem{FPlastina2007}
F. Plastina and T. J. G. Apollaro, Local Control of Entanglement in a Spin Chain, Phys. Rev. Lett. \textbf{99}, 177210 (2007)
%
\bibitem{VAlba2013}
V. Alba, K. Saha, and M. Haque, Bethe ansatz description of edge-localization in the open-boundary XXZ spin chain, J. Stat. Mech. Theor. Exp. \textbf{2013}, P10018 (2013).
%
\bibitem{MHaque2010}
M. Haque, Self-similar spectral structures and edge-locking hierarchy in open-boundary spin chains, Phys. Rev. A \textbf{82}, 012108 (2010).
%
\bibitem{ASharma2014}
A. Sharma and M. Haque, Fine structures in the spectrum of the open-boundary Heisenberg chain at large anisotropies, Phys. Rev. A \textbf{89}, 043608 (2014).
%
\bibitem{GDellaValle2014}
G. Della Valle and S. Longhi, Floquet-Hubbard bound states in the continuum, Phys. Rev. B \textbf{89}, 115118 (2014).
%
\bibitem{HZhong2017}
H. Zhong, Z. Zhou, B. Zhu, Y. Ke, and C. Lee, Floquet Bound States in a Driven Two-Particle Bose-Hubbard Model with an Impurity, Chin. Phys. Lett. \textbf{34}, 070304 (2017).
%
\bibitem{WLiu1911}
W. Liu, Y. Ke, B. Zhu, and C. Lee, Modulation-induced long-range magnon bound states in one-dimensional optical lattices, New J. Phys. \textbf{22}, 093052 (2020).
%
\bibitem{WLiu2019}
W. Liu, Y. Ke, L. Zhang, and C. Lee, Bloch oscillations of multimagnon excitations in a Heisenberg XXZ chain, Phys. Rev. A \textbf{99}, 063614 (2019).
%
\bibitem{VVIvanov2008}
V. V. Ivanov, A. Alberti, M. Schioppo, G. Ferrari, M. Artoni, M. L. Chiofalo and G. M. Tino, Coherent Delocalization of Atomic Wave Packets in Driven Lattice Potentials, Phys. Rev. Lett. \textbf{100}, 043602 (2008).
%
\bibitem{MGTarallo2012}
M. G. Tarallo, A. Alberti, N. Poli, M. L. Chiofalo, F.-Y. Wang and G. M. Tino, Delocalization-enhanced {B}loch oscillations and driven resonant tunneling in optical lattices for precision force measurements, Phys. Rev. A \textbf{86}, 033615 (2012).
%
\bibitem{BZhu2020}
B. Zhu, Y. Ke, W. Liu, Z. Zhou, and H. Zhong, Floquet-surface bound states in the continuum in a resonantly driven one-dimensional tilted defect-free lattice, Phys. Rev. A \textbf{102}, 023303 (2020).
%
\bibitem{MTakahashi1997}
M. Takahashi, Half-filled Hubbard model at low temperature, J. Phys. C: Solid State Phys. \textbf{10}, 1289 (1997).
%
\bibitem{Lee2004}
C. Lee, Bose-Einstein Condensation of Particle-Hole Pairs in Ultracold Fermionic Atoms Trapped within Optical Lattices, Phys. Rev. Lett. \textbf{93}, 120406 (2004).
%
\bibitem{LiZhang2019}
L. Zhang, Y. Ke, and C. Lee, Magnetic phase transitions of insulating spin-orbit coupled Bose atoms in one-dimensional optical lattices, Phys. Rev. B \textbf{100}, 224420 (2019).
%
\bibitem{AWidera2004}
A. Widera, O. Mandel, M. Greiner, S. Kreim, T. W. H\"{a}nsch, and I. Bloch, Entanglement Interferometry for Precision Measurement of Atomic Scattering Properties, Phys. Rev. Lett. \textbf{92}, 160406 (2004).
%
\bibitem{CGross2010}
C. Gross, T. Zibold, E. Nicklas, J. Est\`{e}ve, and M. K. Oberthaler, Nonlinear atom interferometer surpasses classical precision
limit, Nature \textbf{464} 1165 (2010).
%
\bibitem{ALGaunt102004062013}
A. L. Gaunt, T. F. Schmidutz, I. Gotlibovych, R. P.
Smith, and Z. Hadzibabic, Bose-Einstein Condensation of Atoms in a Uniform Potential, Phys. Rev. Lett. \textbf{110}, 200406
(2013).
%
\bibitem{ABuchleitner2003}
A. Buchleitner and A. R. Kolovsky, Interaction-Induced Decoherence of Atomic {B}loch Oscillations, Phys. Rev. Lett. \textbf{91}, 253002 (2003).
%
\bibitem{ARKolovsky2003}
A. Kolovsky and A. Buchleitner, Floquet-{B}loch operator for the {B}ose-{H}ubbard model with static field, Phys. Rev. E \textbf{68}, 056213 (2003).
%
\bibitem{NGoldman2015}
N. Goldman, J. Dalibard, M. Aidelsburger, and N. R. Cooper, Periodically driven quantum matter: The case of resonant modulations, Phys. Rev. A \textbf{91}, 033632 (2015).
%
\bibitem{VVGann367222010}
V. V. Gann and Y. A. Kosevich, Bloch oscillations of spin waves in a nonuniform magnetic field, Low. Temp. Phys. \textbf{36}, 722 (2010).
%
\bibitem{YAKosevich252460022013}
Y. A. Kosevich and V. V. Gann, Magnon localization and Bloch oscillations in finite Heisenberg spin chains in an inhomogeneous magnetic field J. Phys.: Condens. Matter \textbf{25}, 246002 (2013).
%
\bibitem{YKe910534092015}
Y. Ke, X. Qin, H. Zhong, J. Huang, C. He, and C. Lee, Bloch-Landau-Zener dynamics in single-particle Wannier-Zeeman systems, Phys. Rev. A \textbf{91}, 053409 (2015).
%
\bibitem{LWang2017}
L. Wang, N. Liu, S. Chen, and Y. Zhang, Quantum walks in the commensurate off-diagonal Aubry-Andr\'{e}-Harper model, Phys. Rev. A \textbf{95}, 013619 (2017).
%
\bibitem{ILGaranovich2008}
I. L. Garanovich, A. A. Sukhorukov, and Y. S. Kivshar, Defect-Free Surface States in Modulated Photonic Lattices, Phys. Rev. Lett. \textbf{100}, 203904 (2008).
%
\bibitem{AlexSzameit2008}
A. Szameit, I. L. Garanovich, M. Heinrich, A. A. Sukhorukov, F. Dreisow, T. Pertsch,
S. Nolte, A. T\"{u}nnermann, and Y. S. Kivshar, Observation of Defect-Free Surface Modes in Optical Waveguide Arrays, Phys. Rev. Lett. \textbf{101}, 203902 (2008).
%
\bibitem{SLonghi2008}
S. Longhi, Coherent control of tunneling in driven tight-binding chains: Perturbative analysis, Phys. Rev. B \textbf{77}, 195326 (2008).
%
\bibitem{SLonghi2013}
S. Longhi and G. DellaValle, Low-energy doublons in the ac-driven two-species Hubbard model, Phys. Rev. A \textbf{87}, 013634 (2013).
%
\bibitem{GDellaVallePRB2014}
G. DellaValle and S. Longhi, Floquet-Hubbard bound states in the continuum, Phys. Rev. B \textbf{89}, 115118 (2014).
%
\bibitem{LinHuLi2020}
L. H. Li, C. H. Lee, S. Mu, and J. B. Gong, Critical non-Hermitian skin effect, Nat. Commun. \textbf{11}, 1 (2020).
%
\bibitem{ChingHuaLee2020}
C. H. Lee, Many-Body Topological and Skin States without Open Boundaries, arXiv:2006.01182v2 (2020).
%
\bibitem{SBravyi32627932011}
S. Bravyi, D. P. DiVincenzo and D. Loss, Schrieffer-Wolff transformation for quantum many-body systems, Ann. Phys \textbf{326}, 2793 (2011).
%
\bibitem{MTakahashi1012891977}
M. Takahashi, Half-filled Hubbard model at low temperature, J.Phys. C \textbf{10}, 1289 (1977).
%
\bibitem{JHeinrichs521491972}
J. Heinrichs and R. O. Jones, On the existence of Stark ladders in finite crystals, J. Phys. C: Solid State Phys. \textbf{5}, 2149 (1972).
%




\end{thebibliography}

\end{document}